\documentclass[preprint,12pt]{elsarticle}

\usepackage{amssymb}
\usepackage{amsmath}
\usepackage[colorlinks=true,linkcolor=blue,citecolor=blue,urlcolor=blue]{hyperref}

\usepackage{graphicx}
\usepackage{subcaption}
\usepackage{bm}
\usepackage{float}
\usepackage{array}
\usepackage{placeins}

\journal{Applied Energy}

\begin{document}

\begin{frontmatter}



\title{A Two-Step Spatio-Temporal Framework for Turbine-Height Wind Estimation at Unmonitored Sites from Sparse Meteorological Data} 


\author[aff1]{Eamonn Organ} \cormark[1] \ead{Eamonn.Organ@ul.ie}
\author[aff1]{Maeve Upton}
\author[aff2]{Denis Allard}
\author[aff2]{Lionel Benoit}
\author[aff1]{James Sweeney}


\affiliation[aff1]{
  organization={Department of Mathematics and Statistics, University of Limerick},
  city={Limerick},
  country={Ireland}
}

\affiliation[aff2]{
  organization={Biostatistics and Spatial Processes (BioSP), INRAE},
  city={Avignon},
  postcode={84914},
  country={France}
}

\cortext[1]{Corresponding author}

\begin{abstract}
Accurate estimates of wind speeds at wind turbine hub heights are crucial for both wind resource assessment and day-to-day management of electricity grids with high renewable penetration. In the absence of direct measurements, parametric models are commonly used to extrapolate wind speeds from observed heights to turbine heights. Recent literature has proposed extensions to allow for spatially or temporally varying vertical wind gradients, that is, the rate at which wind speed changes with height. However, these approaches typically assume that reference height and hub height measurements are available at the same locations, which limits their applicability in operational settings where meteorological stations and wind farms are spatially separated.
In this paper, we develop a two-step spatio-temporal framework to estimate turbine height wind speeds using only open-access observations from sparse meteorological stations. First, a non-parametric generalized additive model is trained on reanalysis data to perform vertical height extrapolation. Second, a spatial Gaussian process model interpolates these hub-height estimates to wind farm locations while explicitly propagating uncertainty from the height extrapolation stage.
The proposed framework enables the construction of high-resolution, sub-hourly turbine-height wind speed time series and spatial wind maps using data available in real time, capabilities not provided by existing reanalysis products. We further provide calibrated uncertainty estimates that account for both vertical extrapolation and spatial interpolation errors. The approach is validated using hub-height measurements from seven operational wind farms in Ireland, demonstrating improved accuracy relative to ERA5 reanalysis while relying solely on real-time, open-access data.


\end{abstract}

\begin{graphicalabstract}
\includegraphics[width=\textwidth]{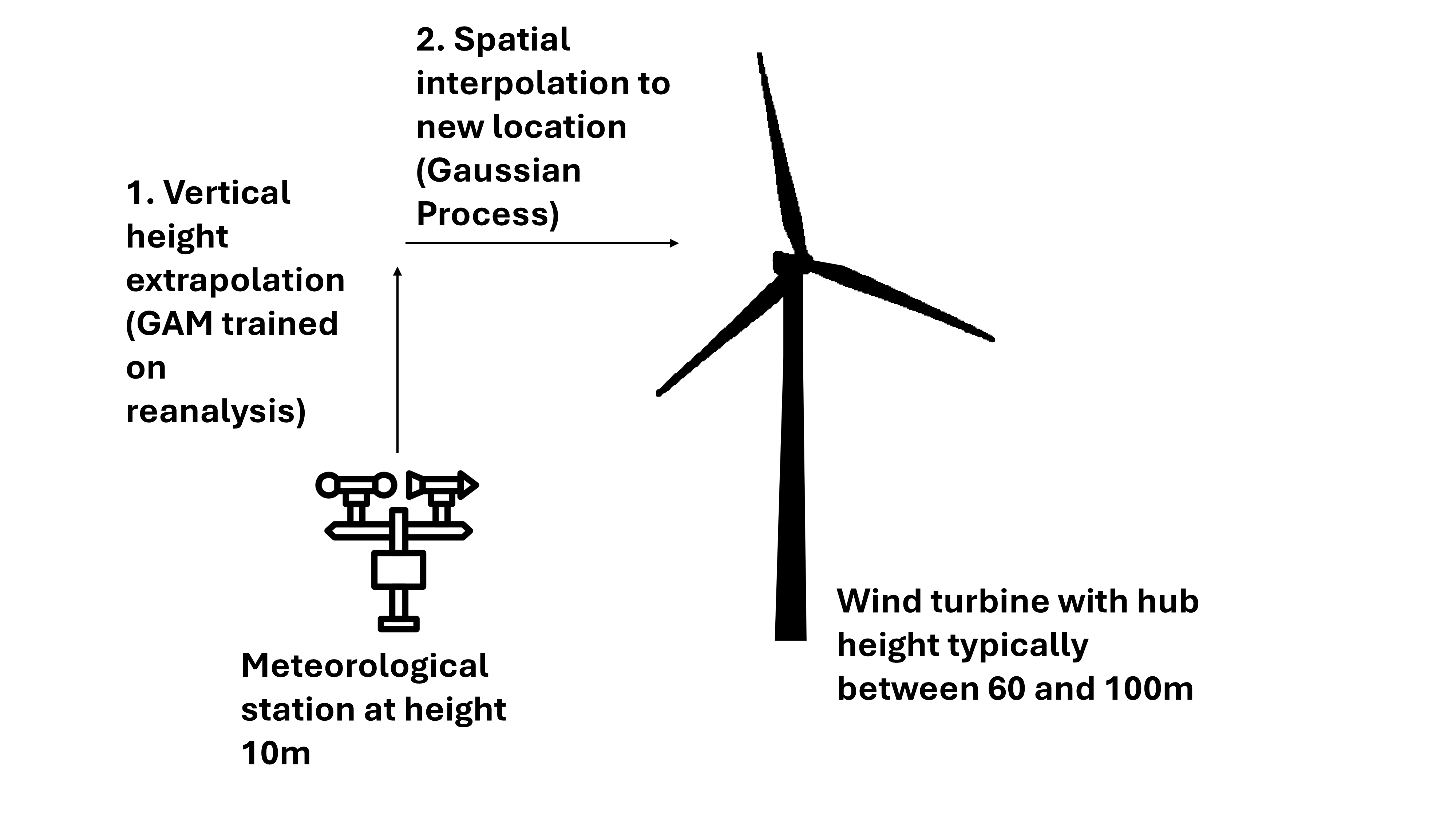}
\end{graphicalabstract}

\begin{highlights}
\item Sub-hourly turbine-height wind speed estimation at unmonitored sites using sparse 10\,m meteorological station data.
\item A two-step framework combining non-parametric height extrapolation and spatial interpolation.
\item Height extrapolation via GAM significantly outperforms parametric shear models.
\item Provides sub-hourly wind speed estimates with quantified uncertainty using data available in real time.
\item Validated at seven operational wind farms and reduces RMSE by 11\% compared to ERA5 reanalysis.
\end{highlights}


\end{frontmatter}



\section{Introduction}

Estimating wind speeds at turbine hub heights is crucial for both wind resource assessment and day-to-day management of electricity grids, and its importance increases as the share of wind power in total generation grows. In Ireland, 34\% of electricity was produced by wind farms in 2024, with the Government setting a target of 80\% by 2030 \cite{GovCAP2024}. To meet that target, the Government has set a goal of 9\,GW of onshore wind energy to be installed by 2030. Currently there is approximately 300 wind farms which provide 4.8\,GW of installed capacity. Based on the current average wind farm size, this would require the construction of approximately 260 new wind farms in the coming years \citep{WindFarmsIRL}.

Wind resource assessment requires knowing wind speed at turbine hub height (i.e., the height at the centre of the rotor) \citep{murthy2017comprehensive}, but official measurements of wind speeds are generally available only at 10\,m (i.e., the standard height of weather stations) \citep{oke2018guide}, leading energy providers to carry out their own surveys at potential sites \cite{weekes2014low}. On-site assessments required for investors can cost from €10,000 to in excess of €100,000, so improved preliminary site screening can lead to targeting sites with higher potential \citep{bailey1997wind,TeagascWind}. Such costs can be prohibitive for small-scale or community energy projects, effectively limiting their ability to undertake rigorous resource assessments and delaying or preventing otherwise viable developments. As a result, reanalysis datasets are popular products for historical wind resource assessment, as they provide wind speeds on a regular grid, typically over several decades \citep{doddy2021reanalysis}. However, these are often coarse in space (with grid sizes from 1 to in excess of 50\,km) and time (typically on an hourly resolution) and released with delays that make them unsuitable for real-time operational use \citep{ramon2019global}.

For day-to-day grid operations, accurate and real-time information on wind conditions is critical in energy systems with high renewable penetration. In particular, errors in wind power forecasts impose substantial operational costs. In the Irish context, McGarrigle and Leahy \citep{mc2015quantifying} estimate that a one percentage point reduction in forecast error could yield annual savings of approximately €2.5 million. While forecasting is not the focus of this study, accurate real-time estimates of wind speed form a key input to short-term forecasting and grid management workflows. For such real-time estimation tasks, statistical methods have been shown to improve upon pure numerical weather prediction outputs \citep{sweeney2020future}. As for wind resource assessment, the real-time estimation tasks involved in day-to-day grid operations require wind speed data at turbine height. 

In the absence of wind measurements at turbine hub height, a common alternative is to vertically extrapolate wind speeds from observations at 10\,m or other near-surface levels. There is extensive research on methods for extrapolating wind speeds from a reference height to turbine hub height, most of which rely on simple parametric formulations, primarily the log law and the power law \citep{houndekindo2025machine}. Gualtieri \cite{gualtieri2019comprehensive} reviews the classical models used in this context. These laws generally depend on parameters relating to surface roughness or air stability. As these parameters are usually unknown or difficult to estimate, fixed approximations are used. These approximations, however, can introduce systematic biases at some locations and cannot capture temporal changes in the vertical wind speed gradient. To complement parametric formulations, recent research has therefore focused on statistical methods that learn how wind speeds vary with height. Crippa et al. \cite{crippa2021temporal} fitted a model in which the power-law parameters vary over time and depend on weather variables other than wind speed. Other studies have explored non-parametric approaches such as random forests and gradient-boosting models to predict wind speeds at target heights \citep{yu2022transfer}. These methods rely on datasets that provide wind speeds at multiple heights, such as physical weather models or specialised remote-sensing instruments like LiDAR (Light Detection and Ranging) \citep{vassallo2020decreasing}. Model training and testing are carried out on multi-height datasets, and these studies therefore assume that wind speeds at several heights are co-located and that wind speeds at a reference height are available at the prediction site. This is a strong practical limitation because in most real-world settings the location where hub-height winds are required is different from where open-access 10\,m measurements are observed, and physical weather model outputs are often too coarse to provide reliable site-specific vertical wind gradients~\citep{gualtieri2022analysing}.

As a result, existing height-extrapolation models cannot be directly applied when hub-height winds are required at locations where only near-surface observations from distant stations are available, which is the most common operational scenario for both grid operators and early-stage site assessment. Moreover, existing approaches do not propagate height-extrapolation uncertainty into subsequent spatial prediction. Consequently, there is currently no established framework capable of estimating hub-height wind speeds at unmonitored locations using only near-surface observations from distant stations, while explicitly propagating height-extrapolation uncertainty into subsequent spatial predictions.

In this context, the present paper addresses a practical but underexplored problem in wind resource modelling: estimating turbine hub-height wind speeds at locations where no on-site measurements exist. This is done using only 10\,m-high wind observations recorded at distant sites. We propose a modelling framework that integrates height extrapolation and spatial interpolation within a unified probabilistic framework, allowing uncertainty from the vertical extrapolation stage to be explicitly propagated into spatial prediction at new locations. This enables hub-height winds to be estimated at unobserved sites using sparse, 10\,m high observations from disparate locations. 

The proposed methodology consists of two linked stages: (i) non-parametric vertical extrapolation of 10 m wind observations to turbine hub heights, and (ii) spatial interpolation of these hub-height estimates to unobserved locations using Gaussian process regression. We propose performing vertical height extrapolation before spatial interpolation, arguing that wind speeds are spatially smoother at hub heights. For vertical height extrapolation, we employ a non-parametric model trained on reanalysis data and then applied to meteorological stations. The resulting hub-height wind speeds are subsequently interpolated using Gaussian process (GP) regression to generate predictions at arbitrary locations. Note that although this study focuses on meteorological stations as reference wind speeds, the proposed methodology can also be applied to extrapolate turbine height wind speeds derived from numerical weather predictions or satellites \citep{verhoef2012high}, which are typically at 10\,m. 

A key contribution of the proposed framework is that it relies exclusively on open-access data streams and can therefore generate turbine-height wind estimates and spatial maps in real time, unlike reanalysis products which are released with multi-day delays. This makes the approach suitable for day-to-day management of renewable grids with high penetrations of wind energy. It can also be applied retrospectively to construct high-resolution hub-height wind-speed time series at prospective sites. Unlike reanalysis datasets, meteorological station observations are available at sub-hourly (often minute-level) resolution, enabling resource assessments at a temporal resolution consistent with the nonlinear power curve of modern turbines. Recent studies have shown that higher temporal resolution wind-speed data are important for accurate wind resource and power estimation, due to the cubic relationship between wind speed and power generation \citep{vest2025apparent}. Another advantage of the framework is its ability to provide full uncertainty quantification for all predictions. This is in contrast with reanalysis data, which provide only point estimates.

To the authors’ knowledge, this is the first framework that jointly performs height extrapolation and spatial interpolation of wind speeds while propagating uncertainty across both stages using only open-access, real-time data. This enables, for the first time, the generation of prospective hub-height wind estimates and minute-level time series at arbitrary locations using only publicly available data, together with calibrated uncertainty estimates.

The layout of the paper is as follows: In section 2 we introduce the datasets used. Section 3 briefly reviews classical vertical height extrapolation approaches, and introduces our non-parametric strategy. Section 4 reviews the spatial models used. Section 5 presents our results on both reanalysis data, and proprietary wind farm data. Finally, Section 6 discusses our conclusions future research directions.

\section{Data}
\label{sec:Data}
This study integrates three different data sources to develop and validate the proposed model. First, meteorological station data provide high-quality, real-time measurements of wind speed at a 10\,m reference height. Second, reanalysis data, the output of physical weather models, provide wind speeds at 10\,m and multiple heights above ground level, which are used to train the height extrapolation model in Section~\ref{sec:Height}. Finally, wind speed measurements from wind farms serve as an external validation dataset to evaluate the accuracy of the complete modelling pipeline.

\subsection{Weather Station Data}

Ground-level wind observations are sourced from the network of 23 meteorological stations operated by Met Éireann, Ireland's national meteorological service \citep{MetEireannWind}. These stations adhere to World Meteorological Organization standards \citep{oke2018guide} and record wind speed at a height of 10\,m.

Of the 23 stations, 18 provide data at a one-minute temporal resolution, while the remaining 5 report wind speeds at an hourly resolution. To ensure temporal consistency with the wind farm dataset used for validation (Section \ref{Sec:WindFarms}), all station data are processed to a 10-minute resolution. For the 18 high-resolution stations, data are averaged over 10-minute intervals, while for the 5 hourly stations, a linear interpolation is applied to generate 10-minute estimates.

\begin{figure}[!htbp]
\centering
\includegraphics[width=0.7\textwidth]{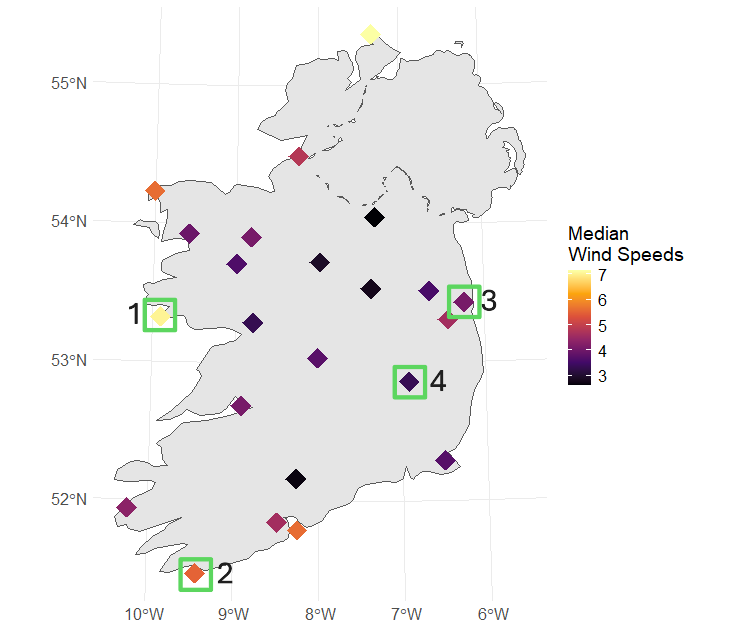}
 \caption{\label{fig:Stationlocs} The location of Ireland's 23 meteorological stations. Highest wind speeds are generally seen along the west coast.}
\end{figure}

\begin{figure}[!htbp]
\centering
\includegraphics[width=0.7\textwidth]{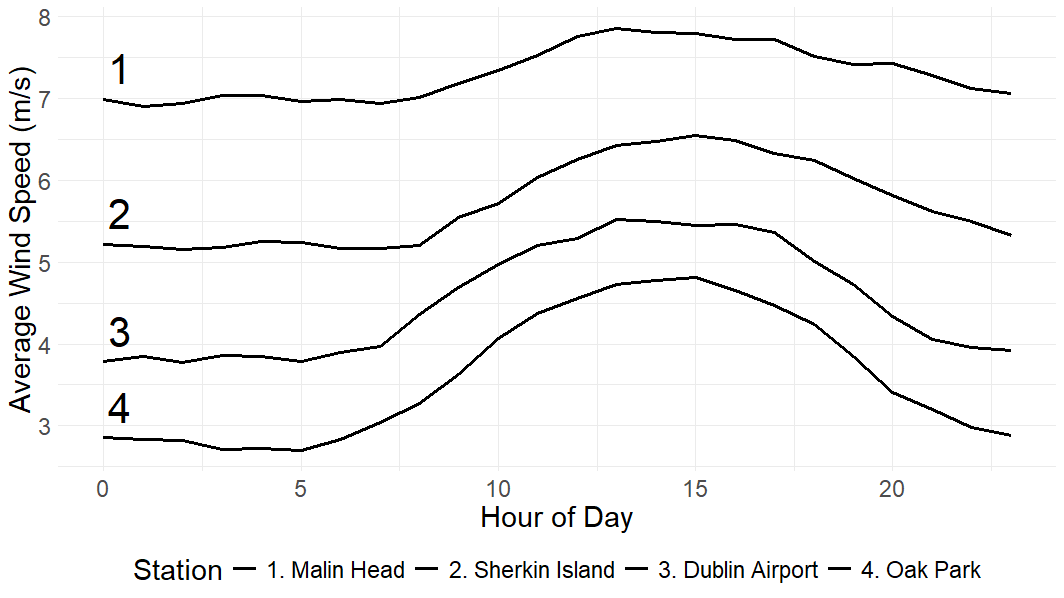}
 \caption{\label{fig:HourlyPattern} Comparison of hourly average speeds at four locations (highlighted in Figure \ref{fig:Stationlocs}. All stations share a similar diurnal (daily) trend).}
\end{figure}

The geographical distribution of the stations, with the colour corresponding to median wind speeds over the year 2023, is shown in Figure \ref{fig:Stationlocs}, illustrating how coastal locations have typically higher wind speeds. Figure \ref{fig:HourlyPattern} depicts the daily cycle of average hourly wind speed at a sample of four locations. This shows a clear diurnal pattern of wind speed, a phenomenon driven by the differential heating of the Earth's atmosphere \citep{dai1999diurnal}, which is consistently captured across stations.

\FloatBarrier
\subsection{Reanalysis Data}
Reanalysis provides a gridded, historical record of the Earth's climate system by combining a physical weather model with observational data via data assimilation \citep{lahoz2014data}. Running the model retrospectively with this constraint produces a more accurate and coherent dataset than the model or observations could provide alone~\citep{lahoz2014data}. These products therefore offer a valuable source of information on numerous weather variables but are inherently delayed, being released days or weeks after the fact, and are thus unsuitable for real-time applications \citep{organ2026enhancing}.

\begin{table}[htbp]
\centering
\caption{Summary of reanalysis and wind atlas datasets used in this study.}
\label{tab:reanalysis_summary}
\setlength{\tabcolsep}{2pt}
\begin{tabular}{p{1.5cm} p{2.5cm} p{3cm} p{2.5cm} p{4.0cm}}
\hline
\textbf{Dataset} &
\textbf{Spatial resolution} &
\textbf{Heights} &
\textbf{Temporal information} &
\textbf{Role in this study} \\
\hline

ERA5 &
$0.25^\circ \times 0.25^\circ$ ($\sim$30 km) &
Wind speeds at \ 10\,m and 100\,m &
Hourly time series (1940 - near present) &
Benchmark for hub-height wind speed prediction \\

NEWA &
3 km &
Wind speeds at 10, 50, 75, 100\,m &
30-min time series (1989-2018) &
Training data for height extrapolation model \\

GWA &
250 m &
Weibull parameters and long-term mean wind speeds at 10, 50 and 100\,m&
Mean and Weibull parameters only (no time series) &
Statistical downscaling of NEWA to station locations and mean covariate \\

\hline
\end{tabular}
\end{table}

\subsubsection{ERA5}
Widely used global reanalysis products for wind applications include ERA5, produced by the European Centre for Medium-Range Weather Forecasts (ECMWF) \citep{hersbach2020era5}, and NASA's Modern-Era Retrospective analysis for Research and Applications, Version 2 (MERRA2) \citep{gelaro2017modern}. These datasets have become a cornerstone for renewable energy resource assessment in numerous studies \citep{doddy2021reanalysis, olauson2018era5}. In particular, Doddy et al.\ \citep{doddy2021reanalysis} identify ERA5 as the most reliable global reanalysis product for wind resource assessment over Ireland, and we therefore adopt ERA5 as a benchmark for comparison in this study. A primary limitation, however, is their relatively coarse resolution. For instance, the ERA5 data used in this study as a benchmark are provided on a $0.25^{\circ} \times 0.25^{\circ}$ grid, translating to a nominal resolution of approximately 28\,km east-west and 17\,km north-south at Irish latitudes.
\par

\subsubsection{New European Wind Atlas (NEWA)}

To train the height extrapolation model, we require a proxy dataset containing wind speeds at both the standard reference height of 10\,m and at turbine-relevant hub heights. We therefore utilise reanalysis data for this purpose. For applications requiring finer spatial scales, several high-resolution reanalysis products have been developed. In this paper, we employ the New European Wind Atlas (NEWA) as our primary data source for model training, which provides a time series of wind speeds spanning almost 30 years \cite{hahmann2020making,dorenkamper2020making}. We use the most recent full year of data available, which is 2018.

The spatial resolution of NEWA is 3\,km, with wind fields available at a temporal resolution of 30 minutes. Wind speeds are provided at several heights, including 10\,m, 50\,m, 75\,m, and 100\,m. Additional heights above 100\,m are also available, but these were excluded from this study as they exceed the range of turbine hub heights in our validation dataset.

\begin{figure}[!htbp]
    \centering
    \begin{subfigure}{0.48\textwidth}
        \centering
        \includegraphics[width=\textwidth]{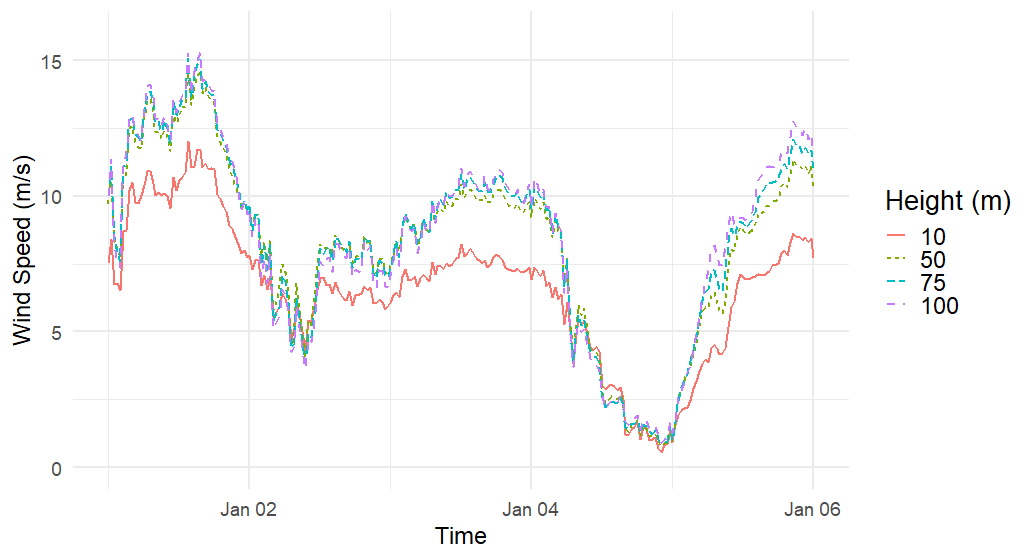}
        \subcaption{1) Malin Head}
    \end{subfigure}
    \hfill
    \begin{subfigure}{0.48\textwidth}
        \centering
        \includegraphics[width=\textwidth]{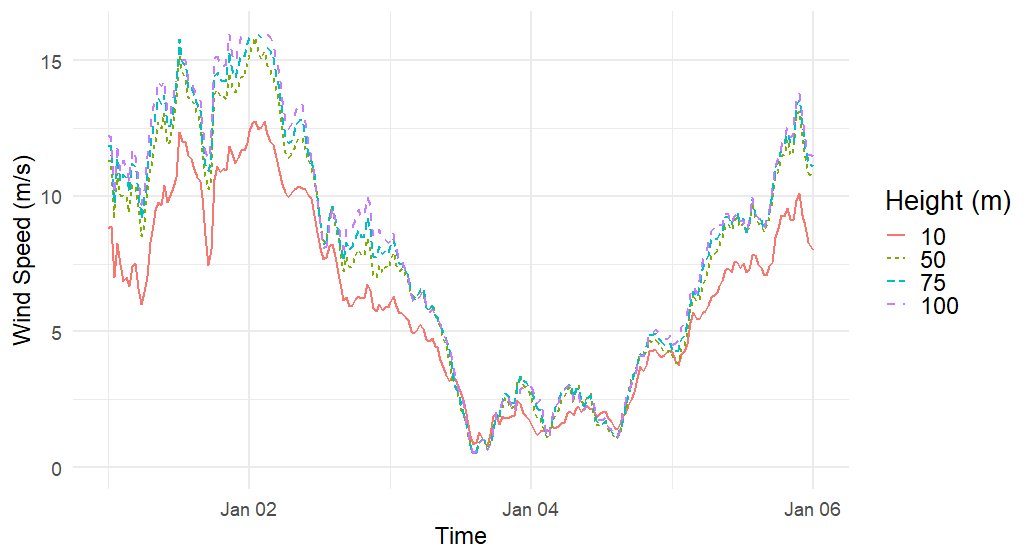}
        \subcaption{2) Sherkin Island}
    \end{subfigure}

    \vspace{0.4cm}

    \begin{subfigure}{0.48\textwidth}
        \centering
        \includegraphics[width=\textwidth]{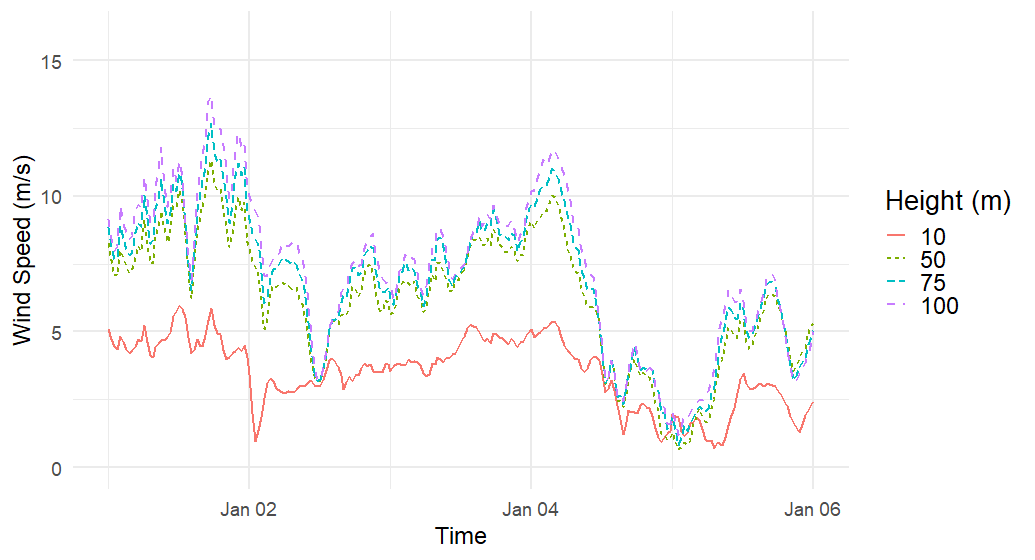}
        \subcaption{3) Dublin Airport}
    \end{subfigure}
    \hfill
    \begin{subfigure}{0.48\textwidth}
        \centering
        \includegraphics[width=\textwidth]{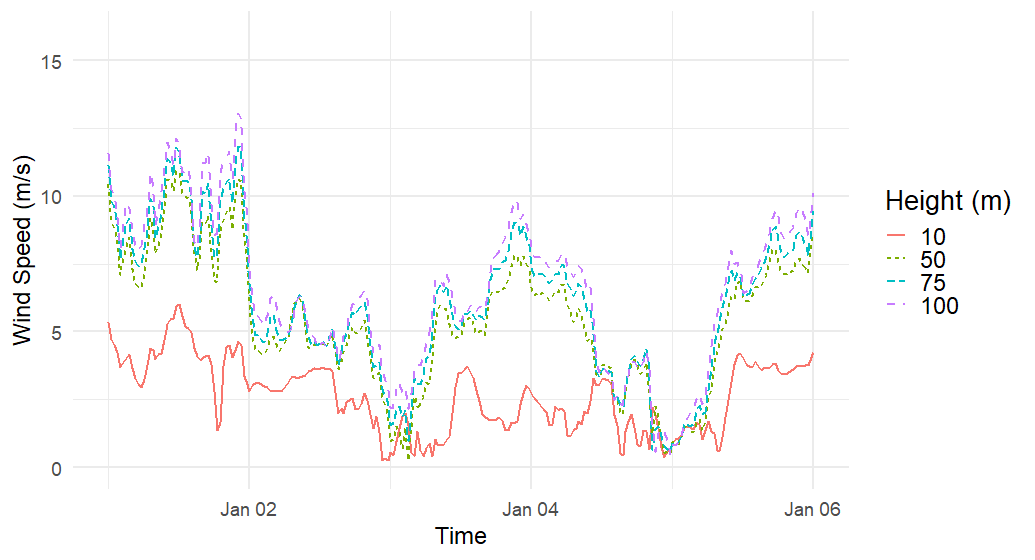}
        \subcaption{4) Oak Park}
    \end{subfigure}

    \caption{\label{fig:wind_heights_4locs}Wind speeds at four locations over the five-day period 1--5 January 2023 (the location of each station is highlighted in Figure \ref{fig:Stationlocs}). Each line represents wind speed at a different height (10\,m, 50\,m, 75\,m, 100\,m ). We choose a sample of four locations that cover both coastal and inland sites. Each line represents wind speed at a different height.}
\end{figure}

\begin{figure}[!htbp]
\centering
\includegraphics[width=0.7\textwidth]{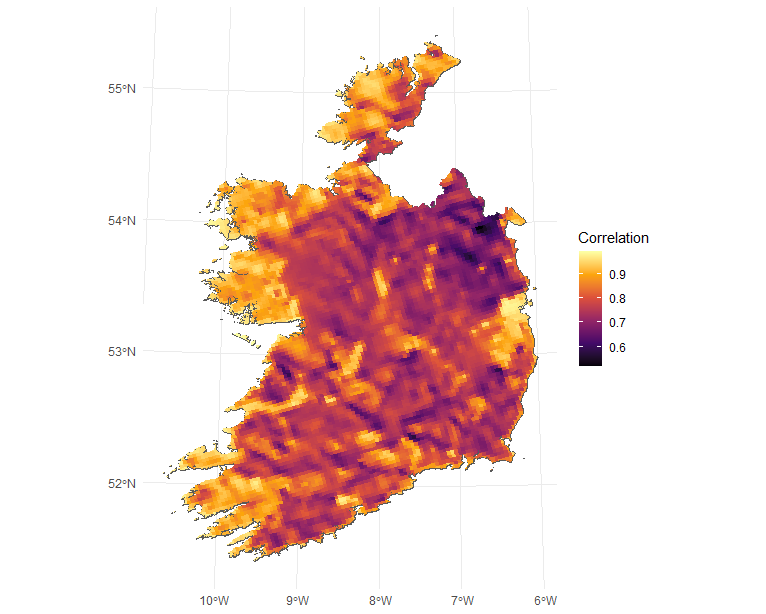}
 \caption{\label{fig:NEWA_Cor} Empirical Pearson correlations between wind speeds at 10\,m and 100\,m. Correlations are generally highest in coastal and upland locations, areas with high surface wind speeds.}
\end{figure}

Figure~\ref{fig:NEWAmean} illustrates the average wind speeds from NEWA over a full year, while Figure~\ref{fig:wind_heights_4locs} shows wind speeds at multiple heights at four sample sites over the five-day period 1--5 January 2023. These plots show that at coastal sites, surface wind speeds (10\,m) closely follow those at hub heights, whereas at inland sites, a greater disparity exists between the two. To summarise this pattern across the whole country, Figure~\ref{fig:NEWA_Cor} shows a map of empirical correlation between 10\,m and 100\,m wind speeds across the NEWA dataset.

\subsubsection{Global Wind Atlas}

While the NEWA provides an extensive time series of wind speeds over several years, the relative coarseness of the grid means the wind speeds at a meteorological station may not closely match the wind speed distribution of the nearest reanalysis grid cell. To address this mismatch, we use a higher-resolution dataset to statistically downscale the wind speed distribution. Specifically, the Global Wind Atlas (GWA) is used \citep{davis2023global}, which has been employed in previous studies to downscale coarser physical models \citep{gruber2022towards,gruber2019assessing}. 

Importantly, the GWA provides only long-term climatological summary statistics rather than a time series of wind speeds, and therefore cannot be used directly for temporal modelling or height extrapolation. Instead, it is used here in complement to NEWA to impose fine-scale spatial structure on the reanalysis time series. The GWA contains summary statistics such as Weibull distribution parameters at a 250\,m resolution, which better capture local topography and surface roughness effects.

\begin{figure}[!htbp]
    \centering

    \begin{subfigure}{0.45\textwidth}
        \includegraphics[width=\linewidth]{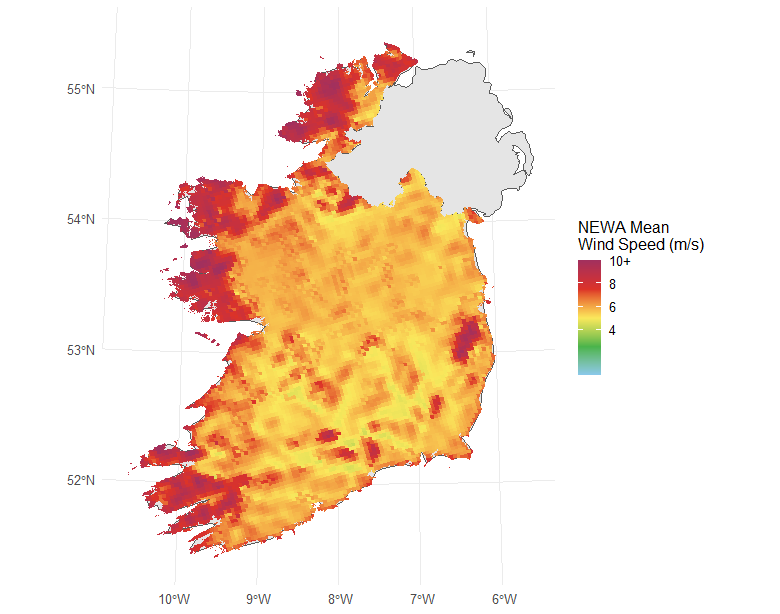}
        \caption{Average 10\,m wind speeds from the New European Wind Atlas (NEWA) - resolution 3 km.}
        \label{fig:NEWAmean}
    \end{subfigure}
    \hfill
    \begin{subfigure}{0.45\textwidth}
        \includegraphics[width=\linewidth]{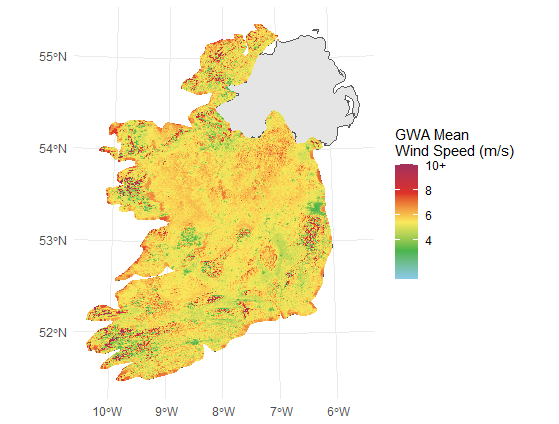}
        \caption{Average 10\,m wind speeds from the Global Wind Atlas (GWA) - resolution 250 m.}
        \label{fig:GWAmean}
    \end{subfigure}
    \caption{\label{fig:Tworeanalysis} Comparison of reanalysis products. The higher resolution of the GWA (250\,m vs. 3\,km) resolves significantly more local topographic variation in wind speed. The NEWA is based on the most recent full year available, 2018. The GWA uses the most recent model update, release 4.0 in June 2025, which provides improved modelling and more recent land use maps \citep{GWADescription4}.}
\end{figure}

Figure~\ref{fig:GWAmean} shows average wind speeds from the GWA across Ireland. When compared to the NEWA plot, the higher spatial resolution reveals additional local variation and fine-scale spatial structure in wind speeds.

\FloatBarrier

\subsection{Wind Farm Data}
\label{Sec:WindFarms}
\begin{figure}[!htbp]
\centering
\includegraphics[width=0.7\textwidth]{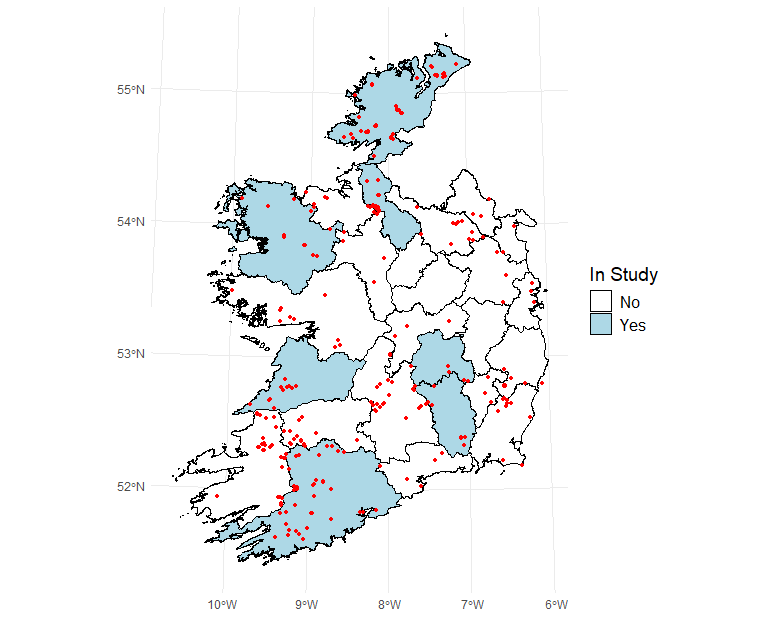}
 \caption{\label{fig:WindFarmloc} The locations of Ireland's wind farms. There are over 300 wind farms in Ireland. The names and locations of wind farms are publicly available and published by the Sustainable Energy Authority of Ireland \citep{WindFarmsIRL,SEAIWindAtlas}. The highlighted counties contain the seven wind farms used for validation. }
 \end{figure}

To illustrate the potential of our framework and assess the accuracy of the proposed models, we use data from a selection of seven wind farms provided by an industry partner. Ireland currently has over 300 operational wind farms \citep{WindFarmsIRL,SEAIWindAtlas}, with their locations shown in Figure~\ref{fig:WindFarmloc}. The counties containing the seven validation farms are highlighted in blue, though specific sites are not identified to maintain data privacy.

Each wind farm comprises of multiple turbines, each recording wind speeds independently. Therefore, model validation can be performed using either the average wind speed across the farm or the speed recorded at an individual turbine. In this study, we use two summary measures for comparison: 
\begin{enumerate}
    \item the highest recorded wind speed across all turbines at each time point, and
    \item the average wind speed across the farm at each time point.
\end{enumerate}

\begin{figure}[!htbp]
\centering
\includegraphics[width=0.7\textwidth]{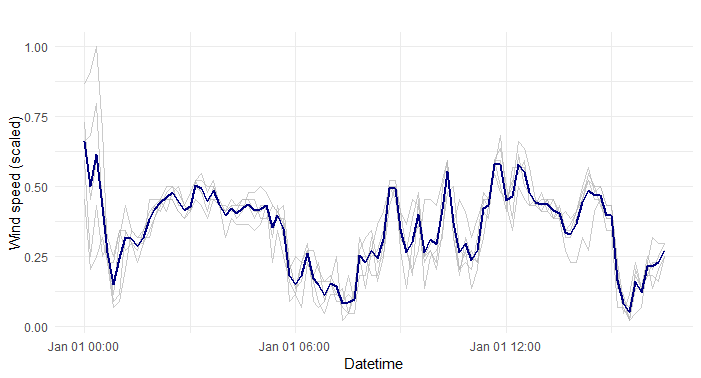}
 \caption{\label{fig:SingleFarmSpeeds} Individual turbine scaled wind speeds (grey) and average wind speed (navy) at a single wind farm over a five-day period.}
\end{figure}

Wind speeds within a farm are influenced by the wake effect, where downstream turbines experience reduced wind speeds due to the energy extraction of upstream turbines \cite{gonzalez2012wake,adaramola2011experimental}. We therefore interpret the maximum recorded speed as being less affected by wake losses, while the  farm average reflects the aggregate effect of wake losses. Figure \ref{fig:SingleFarmSpeeds} shows the wind speeds at a single farm, comparing both the average and the speeds of each individual turbine. Since reanalysis datasets are designed to represent unobstructed, free-stream wind conditions, we expect our predicted wind speeds to more closely match the maximum recorded speeds, while systematically overestimating the farm average values unless wake losses are explicitly accounted for.

Turbines typically measure wind speeds at their hub, which is the height at the centre of the rotor. Hub heights can vary from 60 to 100\,m for typical wind turbines found in Ireland, depending on the turbine model. Turbine wind speeds are provided at a 10 minute resolution, and all modelling in this study is therefore performed at this time scale.

\FloatBarrier
\section{Modelling Framework}

The objective of our modelling framework is to use weather station observations to predict wind speeds at wind farm locations and turbine hub heights. This introduces two challenges: (i) a spatial interpolation process is required to interpolate wind speeds from meteorological stations to wind farm sites, and (ii) a vertical extrapolation model is needed to predict wind speeds at turbine hub height, conditional on 10m observations, which are the standard reporting height for meteorological stations.  

A practical constraint is that wind speed observations at turbine hub heights are typically proprietary and not publicly accessible. Consequently, stakeholders often have limited or no access to this data. In this study, wind farm hub height measurements are therefore used only as an external test set, while all model training relies exclusively on the publicly available datasets described in Section~2.  

\begin{figure}[!htbp]
\centering
\includegraphics[width=0.7\textwidth]{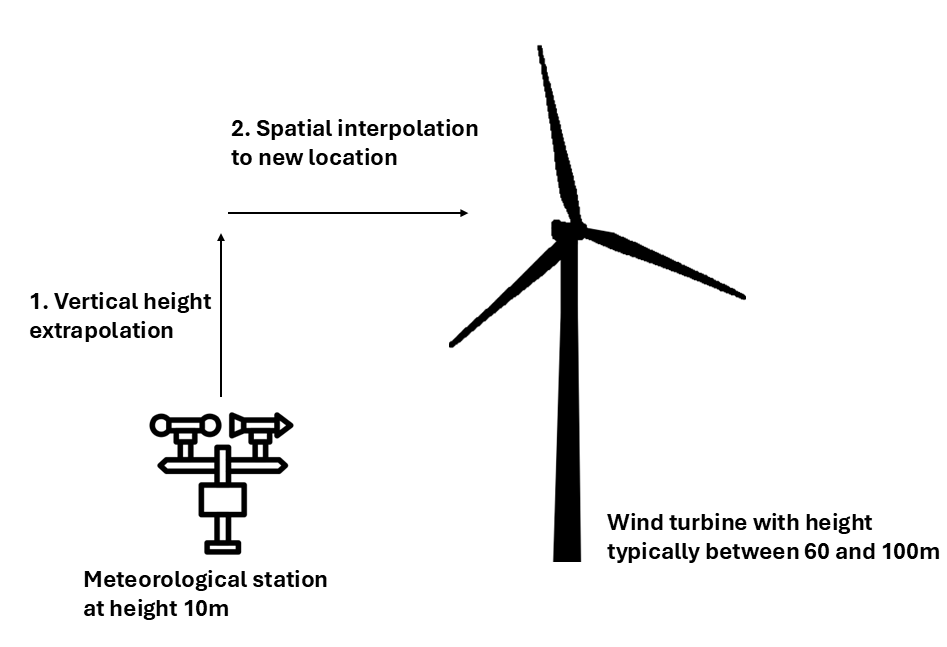}
 \caption{\label{fig:ModelVisual} Schematic of the two-step process: (1) vertical extrapolation from 10\,m to turbine hub height, followed by (2) spatial interpolation from meteorological stations to wind farm locations.}
\end{figure}

Our prediction framework addresses these challenges through a sequential two-step approach, illustrated in Figure~\ref{fig:ModelVisual}:
\begin{enumerate}
    \item \textbf{Vertical height extrapolation:} A generalised additive model (GAM) is trained on reanalysis data to learn vertical wind-speed profiles and to predict turbine hub-height wind speeds from 10\,m reference observations at meteorological stations.
    \item \textbf{Spatial interpolation:} The resulting hub-height wind speed predictions are then interpolated to target wind farm locations using a spatial statistical model
\end{enumerate}

\begin{figure}[!htbp]
\centering

\begin{subfigure}[b]{0.32\textwidth}
    \includegraphics[width=\textwidth]{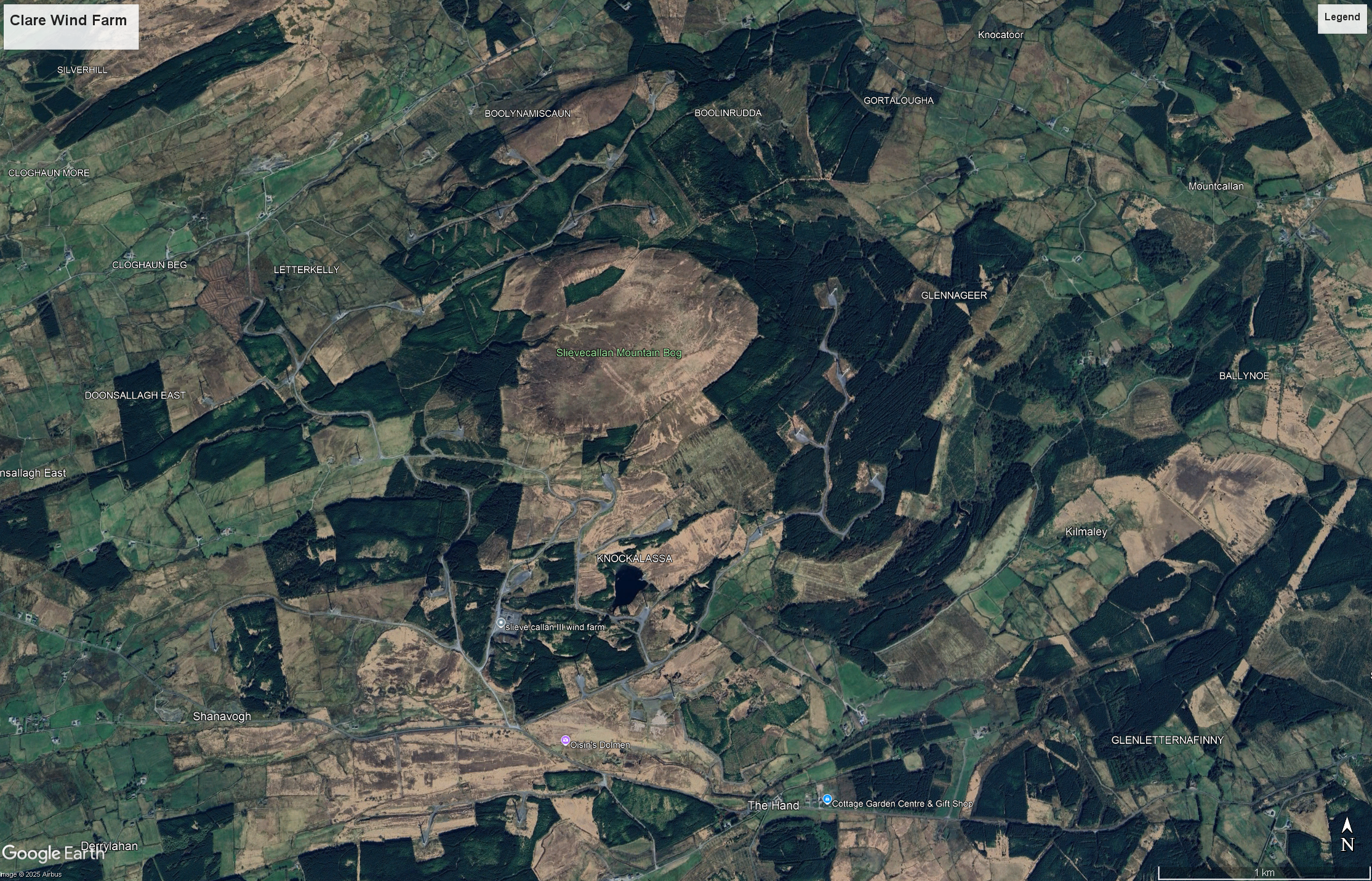}
    \caption{Satellite image of the region. Image Source: Google Earth, Maxar Technologies}
    \label{fig:sat_image}
\end{subfigure}
\hfill
\begin{subfigure}[b]{0.34\textwidth}
    \includegraphics[width=\textwidth]{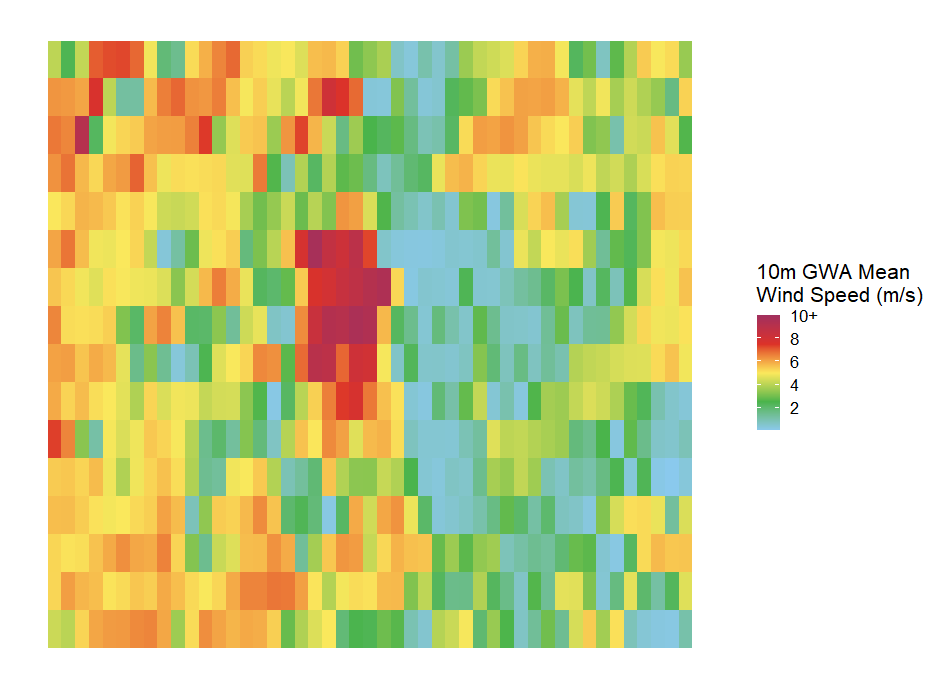}
    \caption{GWA mean wind speed at 10\,m}
    \label{fig:gwa_10m}
\end{subfigure}
\hfill
\begin{subfigure}[b]{0.29\textwidth}
    \includegraphics[width=\textwidth]{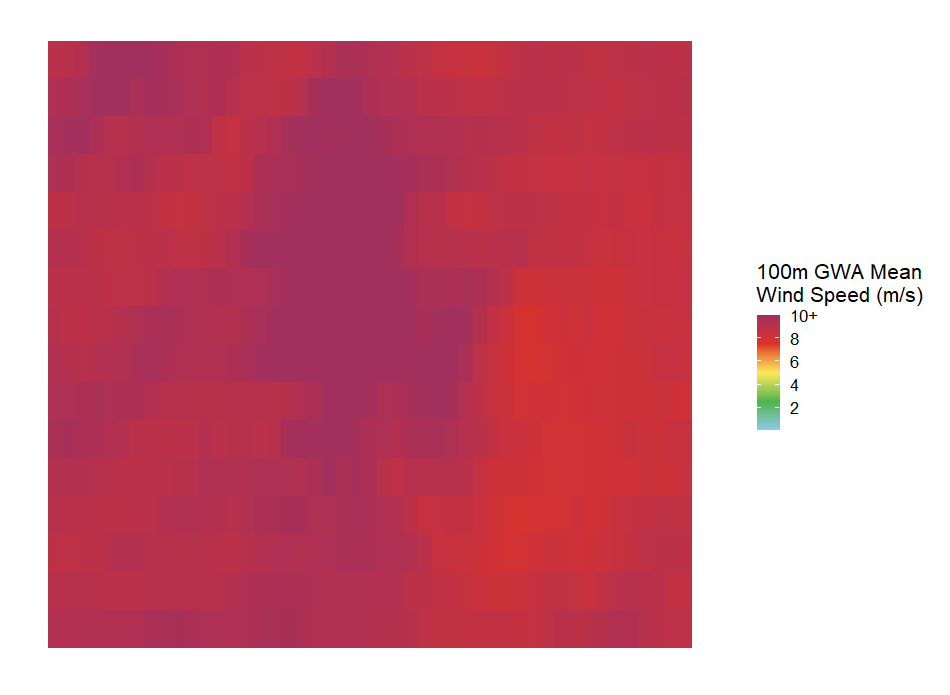}
    \caption{GWA mean wind speed at 100\,m}
    \label{fig:gwa_100m}
\end{subfigure}

\caption{Comparison of wind speeds at a sample wind farm (outside the validation dataset) in the west of Ireland: (a) satellite image. (b) GWA mean wind speed at 10\,m showing fine-scale variability. The average speeds within the image have a range of 9.7\,m/s and a standard deviation of 2.1\,m/s.  (c) GWA mean wind speed at 100\,m showing smoother wind field at turbine height. The average wind speeds have a range of 3.4\,m/s and a standard deviation of 0.65\,m/s. The large amounts of forestry around the turbines cause large variations in average wind speeds at 10\,m, while the 100\,m wind speeds significantly lower levels of variation.}
\label{fig:wind_comparison}
\end{figure}

The order of these steps is motivated by the fundamental meteorological principle that wind fields are smoother and exhibit longer-range spatial correlation at higher altitudes, free from the fine-scale turbulence induced by surface roughness \citep{schlez1998horizontal}. As illustrated in Figure~\ref{fig:wind_comparison}, the mean wind field at 100\,m is considerably smoother than at 10\,m, where local topography and vegetation introduce strong small-scale variability. Performing height extrapolation first leverages this smoother, more spatially coherent field for the subsequent interpolation step, which is likely to improve accuracy.

\subsection{Reanalysis Wind Speeds Preprocessing}
\label{sec:Preprocessing}
Before fitting the height extrapolation models, we preprocess the reanalysis data to address its inherent spatial coarseness. While the New European Wind Atlas (NEWA) provides a high-quality temporal record, its 3\,km resolution may not fully capture the local wind climate at a specific point, such as a meteorological station location. To incorporate finer-scale topographic effects, we statistically downscale the NEWA data using the high-resolution wind climatology from the Global Wind Atlas (GWA).

The downscaling procedure is a quantile-quantile transformation at each of the NEWA height levels\citep{michelangeli2009probabilistic}, which maps the empirical distribution of the NEWA data at a grid point to a theoretical distribution defined by the GWA's parameters at the location of the reference sites. This process effectively imparts the high-resolution spatial structure of the GWA's climatology onto the NEWA time series.

Let \(\mathbf{w}_{\text{NEWA}} = (w_1,\dots,w_n)\) be a time series of wind speeds of size \(n\) from the nearest NEWA grid point to a meteorological station, with empirical cumulative distribution function (CDF) \(F_n(w) = \frac{ \# (w_i \leq w)}{n}\). The empirical cumulative probability (or percentile rank) associated with an observation \(w_i\) is \(p_i = F_n(w_i)\).

We assume that the long-term wind speed distribution at a location $s$ follows a parametric distribution with CDF $\hat{F}_{s}(w; \Theta_s)$, where $\Theta_s$ denotes the distribution parameters. For this purpose, we adopt the two-parameter Weibull distribution, which is widely used to model wind speed climatology \citep{carta2009review, wais2017review, bowden1983weibull}. Its probability density function (PDF) and CDF are given by:
\begin{align}
    f(x; k, \lambda) &= \frac{k}{\lambda}\left(\frac{x}{\lambda}\right)^{k-1} \exp\left[-\left(x/\lambda\right)^k\right], \qquad
    F(x; k, \lambda) = 1 - \exp\left[-\left(x/\lambda\right)^k\right],
\label{eq:WeibullCDF}
\end{align}
where $k>0$ and $\lambda>0$ are the shape and scale parameters, respectively. These parameters are directly available from the Global Wind Atlas (GWA) at 250\,m resolution \citep{davis2023global}, and the Weibull distribution has been shown to accurately represent Irish wind speed distributions \citep{organ2026enhancing}. For each station location, we extract the corresponding GWA parameters $(k_s, \lambda_s)$, define $\hat{F}_s$ using Equation~\eqref{eq:WeibullCDF}, and apply the quantile mapping in Equation~\eqref{eq:qq_transform} to the NEWA time series to obtain the downscaled dataset.

The downscaled wind speed, \(\widetilde{w}_{s,i}\), is obtained by applying the inverse CDF transform using the GWA-derived theoretical distribution:
\begin{equation}
    \widetilde{w}_{s,i} = \hat{F}_{s}^{-1}\left(p_i\right) = \hat{F}_{s}^{-1}\left( F_n(w_i) \right).
\label{eq:qq_transform}
\end{equation}

As the downscaled NEWA dataset only reports wind speeds at one reference height and three target heights (50\,m, 75\,m, 100\,m), there is a risk that models trained directly on these values may overfit to the discrete levels or fail to capture the continuous effect of height on wind speed. To address this, we construct a smooth vertical profile of wind speed at each time and location by interpolating the three target heights with a quadratic curve. From this fitted curve, wind speeds are then sampled at 5 m intervals, producing a denser set of pseudo-observations. A 5 m spacing was chosen as a practical balance: it provides sufficient vertical resolution to train a non-parametric extrapolation model while avoiding unnecessary computational overhead.


\subsection{Height Extrapolation}
\label{sec:Height}

Wind speed extrapolation from observations at a reference height (typically 10\,m) to turbine hub height is traditionally performed using parametric models such as the power law or logarithmic law. A comprehensive overview is given in \cite{gualtieri2019comprehensive}, which found the power log to be the most popular and generally most reliable. In its generic form, the power law is defined as
\begin{equation*}
    W_{h} = W_{10}\left(\frac{h}{10}\right)^{\alpha},
\end{equation*}
where $W_{h}$ and $W_{10}$ are wind speeds at heights $h$ and 10\,m, respectively, and $\alpha$ is a wind shear parameter dependent on surface roughness and atmospheric stability. Although $\alpha$ is known to vary in space and time, fixed approximations are often used for simplicity, with $\alpha = 1/7$ being the most common choice \cite{houndekindo2025machine,panofsky1984atmospheric}.  

\begin{figure}[!htbp]
\centering
\includegraphics[width=0.7\textwidth]{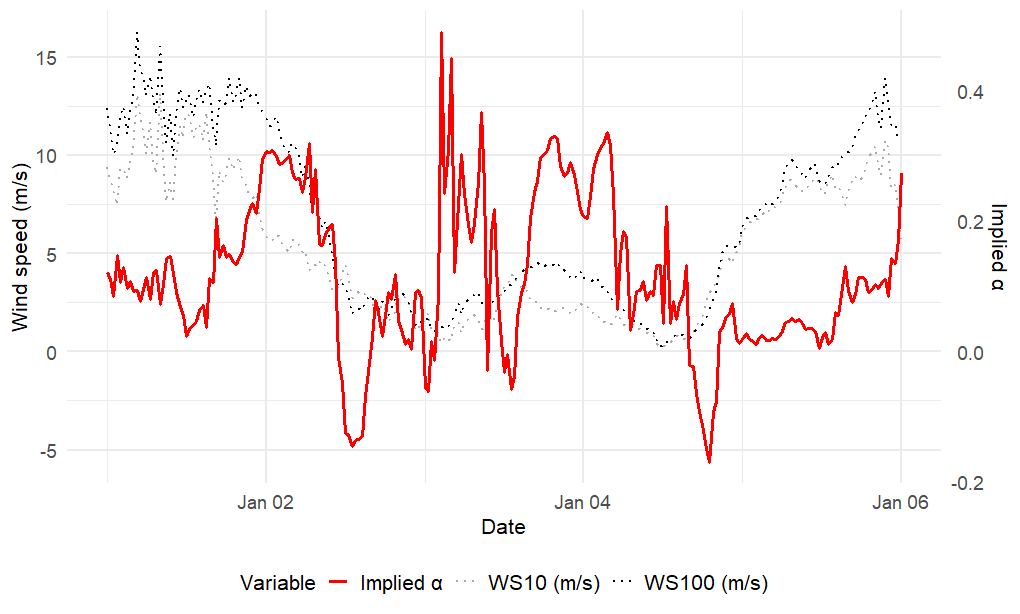}
 \caption{\label{fig:ImpliedAlpha} Five days of 10\,m and 100\,m wind speeds at Malin Head, labelled (1) in Figure \ref{fig:wind_heights_4locs}. The $\alpha$ value at each time point derived from the two wind measurements is plotted. This plot illustrates that $\alpha$ shows a high amount of variability, and constant or periodic harmonics may struggle to fully capture the relationship }
\end{figure}

Extensions have been proposed to account for this variability by allowing \(\alpha\) to depend on covariates. For instance, Crippa et al.\cite{crippa2021temporal} modelled \(\alpha(t)\) as a function of diurnal harmonics, with a separate model fit to each location in their dataset:
\begin{equation}
    \alpha(t) = \alpha_0 + \sum_{i=1}^{P} \beta_{i}\sin\!\left(\tfrac{2\pi ti}{24}\right) + \beta_{i}^{'}\cos\!\left(\tfrac{2\pi ti}{24}\right),
\end{equation}
where $\alpha_{0}$ is a constant mean term, and $\beta_{i}$ and $\beta_{i}^{'}$ are the coefficients for $P$ periodic harmonics. 
Such approaches substantially improved predictions relative to fixed \(\alpha\) values. However, our analysis of reanalysis data indicates that empirically derived \(\alpha\) values display high variability and strong, non-linear interactions with surface wind speed (Figure~\ref{fig:ImpliedAlpha}), limiting the effectiveness of parametric approaches that rely on a simplified structural form. The high variability suggests that direct non-parametric modelling of wind speeds may be preferable.  

Previous studies using non-parametric approaches also provide good accuracy for wind height extrapolation predictions \citep{yu2022transfer,vassallo2020decreasing}.
We choose to adopt a flexible, non-parametric approach using a Generalized Additive Model (GAM) \cite{wood2017generalized}. GAMs have been shown to achieve accuracy comparable to machine learning methods, but offer parsimonious inference and interpretable outputs \citep{doohan2026comparison}.   
In our framework the GAM predicts the \textit{square root} of hub-height wind speeds in the processed NEWA dataset. This transformation is commonly used to reduce positive skewness and to approximate Gaussianity in wind-speed data \citep{lenzi2020spatiotemporal}. As described in Section~\ref{sec:Preprocessing}, the NEWA time series are first statistically downscaled using Weibull distributions derived from the Global Wind Atlas in order to encode fine scale spatial structure. 
For Irish wind climates, the Weibull shape parameter is typically close to 2 \citep{organ2026enhancing}. If a wind speed follows a Weibull distribution with shape parameter $k$ and scale parameter $\lambda$, then its square root also follows a Weibull distribution with shape parameter $2k$ and scale parameter $\sqrt{\lambda}$. Consequently, the square-root transformation yields a distribution with shape parameter close to 4, a regime in which the Weibull distribution exhibits low skewness and is well approximated by a Gaussian distribution \citep{cui2003some}. This motivates the use of a Gaussian likelihood on the square-root-transformed scale, which simplifies model fitting while remaining consistent with the empirical distribution of Irish wind speeds.

Predictors include the square root of 10\,m wind speed, turbine hub height, diurnal cycles, and wind direction components. These covariates are chosen as crucially, they are all available in real time at meteorological stations. The model is specified as:
\begin{equation}
\sqrt{W_{h}} = \beta_{0} +
s_{1}(\sqrt{W_{10}}) + s_{2}(h) + s_{3}(h, W_{10})
+ \beta_{1}\sin\!\left(\tfrac{2\pi t}{24}\right)
+ \beta_{2}\cos\!\left(\tfrac{2\pi t}{24}\right)
+ \beta_{3}U + \beta_{4}V,
\end{equation}
where $\beta_0$ is an intercept term, and $\beta_1, \dots, \beta_4$ are coefficients associated with the linear covariates. $W_h$ and $W_{10}$ denote wind speeds at hub height $h$ and 10\,m, respectively; $t$ denotes time in hours; $s(\cdot)$ denotes generic smooth functions; and $U$ and $V$ are the east-west and north-south components of the unit wind-direction vector at 10\,m. All smooth terms use the default spline bases in \texttt{mgcv} (thin-plate regression splines), with smoothing parameters estimated via restricted maximum likelihood (REML). Basis dimensions were checked using standard \texttt{mgcv} diagnostics. $s_{2}$ and $s_{3}$ were left at their default values, while for $s_{1}$ the basis dimension was increased until the k-index was close to 1\citep{wood2017generalized}. 

A separate model is fit for each meteorological station using reanalysis wind speeds extracted from the nearest NEWA grid point and statistically downscaled using the Global Wind Atlas, as described in Section~\ref{sec:Preprocessing}, accounting for local effects such as surface roughness. Each training observation corresponds to a specific time point and height and are treated as independent observations. While this formulation ignores temporal and vertical dependence in the reanalysis data, it is consistent with the intended operational use of the model, where predictions at any given time must rely solely on the 10\,m wind observation available at that location. This setup ensures that model training and evaluation reflect the information constraints present in real-time applications. Implementation uses the \texttt{mgcv} package in R, with efficient estimation (typically under 10 seconds per model on a standard laptop). The process is easily parallelised if fitting is required for many locations.

\begin{figure}[!htbp]
\centering
\includegraphics[width=0.7\textwidth]{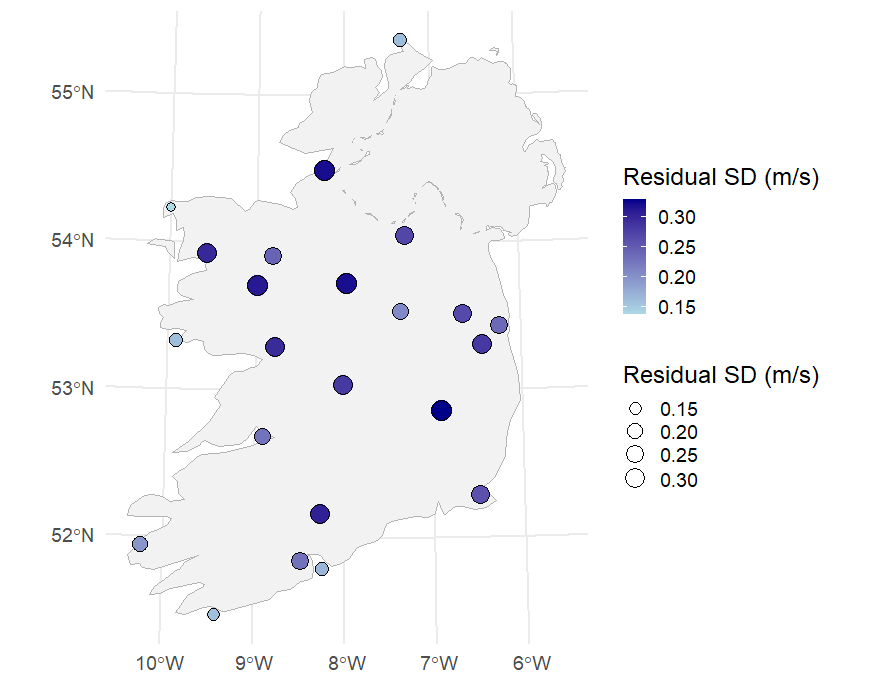}
 \caption{\label{fig:NEWA_GAM_variances} Estimated standard deviation $\sigma_{s}$ of the height extrapolation model at each wind farm location.}
\end{figure}

Model uncertainty is quantified directly from the fitted GAMs. For each station specific GAM, residual variance is estimated under the assumption of Gaussian errors with constant variance on the square root scale. This residual variance is used as a measure of uncertainty in the height-extrapolated wind speeds at each station. Figure~\ref{fig:NEWA_GAM_variances} shows the resulting station-specific variance estimates.

Uncertainty is generally higher at inland sites, while coastal stations—despite higher mean wind speeds—exhibit lower extrapolation uncertainty, possibly due to smoother vertical wind profiles near open seas. As seen in Figure \ref{fig:wind_heights_4locs} and Figure \ref{fig:NEWA_Cor}, at coastal location, wind speeds at 10\,m more closely follow the pattern shown at turbine height.

\subsection{Spatial Interpolation}
\label{sec:Spatial}

To interpolate wind speeds from observed to unobserved locations, we model wind speed as a Gaussian Process (GP) \citep{banerjee2003hierarchical,williams2006gaussian}. Spatial interpolation is performed at turbine hub height, as wind speeds at greater heights are generally smoother and less affected by surface obstacles. Following Section \ref{sec:Height}, spatial inference is performed using the square root of wind speeds, which is approximately Gaussian.

The mean structure of the spatial process incorporates information from the Global Wind Atlas (GWA), with the GWA mean square-root wind speed at location $s$ and at the target height $h$, denoted $\mu_s^{h}$, used as a covariate. The mean square-root wind speed is derived from the GWA Weibull parameters \citep{organ2026enhancing}. As the GWA is available only at fixed reference heights, which generally differ from turbine hub heights, a power-law relationship is used to interpolate the mean wind speed to the target height $h$. This provides a spatially varying baseline that reflects long-term climatological differences between locations. While other spatial covariates such as elevation and distance to the coast have been used in wind speed studies, the GWA mean implicitly incorporates these effects through the physical modelling used to generate the atlas.

The dataset comprises Met Éireann wind speed observations recorded at 10 minute intervals over the full year 2023 (52,560 time points). Fitting a fully spatio-temporal Gaussian process (GP) model to data of this size would be computationally demanding, as exact GP inference scales as $\mathcal{O}(n^{3})$. While scalable approximations such as composite likelihood or Vecchia-type methods could be considered, incorporating an explicit temporal dependence structure is unlikely to substantially improve spatial interpolation in this setting, since observations are available at the same fixed locations for all time points.

We therefore assume temporal independence and treat each time point as an independent realization of the spatial process, using temporal replication to improve estimation of the spatial covariance structure. This approach substantially reduces computational cost while retaining the information needed for accurate spatial interpolation at each time step.
To allow for seasonal variation in both the mean structure and spatial covariance, a separate spatial model is fitted for each calendar month, using the full 10 minute time series within that month for inference and prediction. This monthly modelling strategy reduces computational cost and enables inference to be performed on standard hardware.

The hub-height wind speeds used as inputs to the spatial model are not direct observations but estimates obtained from the height-extrapolation step in Section \ref{sec:Height}. To account for this additional source of uncertainty, we treat the square-root-transformed wind speeds as noisy observations of an underlying latent Gaussian process, with an observation-error variance informed by the height-extrapolation model. As discussed in Section \ref{sec:Height}, the square root transform provides a suitable Gaussian approximation for wind speeds, which allows for efficient spatial inference.  

The square root of the hub-height wind speeds is assumed to follow a multivariate normal (MVN) distribution with mean given by a latent spatial process and covariance equal to the sum of the GAM-based uncertainty and an additional noise term to account for small scale noise at individual meteorological stations:

\begin{equation*}
    \sqrt{W^{h}} \sim \textbf{MVN}\left(\mathbf{f}^{h}, \Sigma_{\text{GAM}} + \sigma^{2}_{\epsilon} I_{n}\right),
\end{equation*}
where MVN denotes the multivariate normal distribution, $n$ is the number of meteorological stations, and $I_n$ is the $n \times n$ identity matrix. The latent process $\mathbf{f}^{h}$ is defined as a Gaussian Process,

\begin{equation*}
    f^{h} \sim GP\left(\mu^{h}, \Sigma_{f}\right),
\end{equation*}
and $\mu_{s}^{h} = \beta_{0} + \beta_{1}\mu_{\text{GWA}}^{h}$ defines the mean structure.

We assume independent measurement errors at each location, making $\Sigma_{GAM} + \sigma^{2}_{\epsilon}I$ diagonal. Unlike the standard homoscedastic assumption (equal variance at all sites), we allow for heteroscedasticity by setting each diagonal entry of $\Sigma_{GAM}$ to $\sigma_{s_i}^{2}$, the estimated variance of the residuals from the GAM predictions at location $s_i$. This provides a location-specific measure of uncertainty. As reanalysis data are often over-smoothed, an additional common nugget term $\sigma_{\epsilon}^{2} I$ is added to the diagonal to capture unresolved microscale variability, which is learned from the data.

Assuming independence between the latent process and measurement error, the marginal distribution of the observed data for each becomes:
\begin{equation*}
    \sqrt{W^{h}} \sim MVN\left(\mu^{h}, \Sigma_{f} + \Sigma_{GAM} +\sigma^{2}_{\epsilon}I\right).
\end{equation*}

The latent GP employs a Matérn covariance function, which is widely used for spatial processes \citep{banerjee2003hierarchical}. To simplify inference and avoid identifiability issues between range and smoothness parameters, we fix the smoothness parameter $\nu = 1$, yielding:
\begin{equation*}
    \Sigma_{f,ij} = \text{cov}(s_{i}, s_{j}) = \kappa d K_1\left(\kappa d\right),
\end{equation*}
where $d$ denotes the Euclidean distance between sites $s_i$ and $s_j$, and $K_{\nu}$ is the modified Bessel function of the second kind with parameter $\nu$. $\kappa$ is a scale parameter, which smaller values of $\kappa$ denote a longer spatial correlation.

For each month, we maximize the log-likelihood with respect to the fixed-effect coefficients and GP hyperparameters:
\begin{equation*}
    \Theta = \{\kappa, \sigma_{f}, \sigma_{\epsilon}, \beta_{0}, \beta_{1}\}.
\end{equation*}
The optimization is carried out in R using the \textit{optim} function with the Broyden–Fletcher–Goldfarb–Shanno (BFGS) algorithm:
\begin{equation*}
    \log L(\mathbf{Y}|\Theta) \propto -\frac{1}{2} \sum_{t=1}^{n_{t}}\left[(\sqrt{W^{h}_{t}} - \boldsymbol{\mu})^{T} \mathbf{C}^{-1} (\sqrt{W^{h}_{t}} - \boldsymbol{\mu}) + \log\det(\mathbf{C})\right],
\end{equation*}
where $C = \Sigma_{f} + \Sigma_{\text{GAM}} + \sigma_{\epsilon}^{2}I$. $n_{t}$ denotes the number of unique 10 minute points in a given month (e.g., $n_{t}=4464$ for a 31 day month).

As multivariate optimisation routines can be sensitive to starting values, initial values for the spatial range and variance parameters were chosen based on estimates reported in previous studies of Irish wind speeds \citep{organ2026enhancing}. Fixed effect coefficients were initialised using ordinary least squares estimates.

Once optimal hyperparameters are obtained from the BFGS algorithm, the posterior mean and variance of the wind speed at wind farm locations can be derived. 
Let $\sqrt{W}$ denote the vector of observed wind speeds at station locations $S$, and $\sqrt{W^{*}}$ the (unobserved) vector of wind speeds at wind farm locations $WF$. 
We define the joint covariance matrix of all locations as
\[
\mathbf{C} =
\begin{bmatrix}
\mathbf{C}_{S,S} & \mathbf{C}_{S,WF} \\
\mathbf{C}_{WF,S} & \mathbf{C}_{WF,WF}
\end{bmatrix},
\]
where each block $\mathbf{C}$ denotes the covariance between locations in sets $S$ and $WF$.

Then, the conditional (posterior) distribution of $\sqrt{W^{*}}$ given $\sqrt{W}$ and parameters $\Theta$ is given by standard GP results \citep{rue2005gaussian}:
\[
\sqrt{W^{*}} \mid \sqrt{W}, \Theta \sim 
\text{MVN}\!\left(
\boldsymbol{\mu}^{*} + 
\mathbf{C}_{WF,S}\mathbf{C}_{S,S}^{-1}(\sqrt{W} - \boldsymbol{\mu}),
\;
\mathbf{C}_{WF,WF} - 
\mathbf{C}_{WF,S}\mathbf{C}_{S,S}^{-1}\mathbf{C}_{S,WF}
\right).
\]
This posterior distribution provides both predicted hub-height wind speeds and their associated uncertainty at each wind farm location.

Taken together, the preprocessing, height extrapolation, and spatial interpolation form an end-to-end framework for predicting turbine hub height wind speeds at unobserved locations. Once trained, the framework can be applied in real time using only 10\,m wind speed and direction observations from meteorological stations. These inputs are first extrapolated to hub height and then spatially interpolated to target locations, producing turbine-height wind speed predictions without requiring access to turbine measurements or reanalysis data.

\section{Results}
\label{sec:Results}
The results are presented in two parts. First, we validate the height extrapolation model (Section~\ref{sec:Height}) using a held-out test set from the reanalysis data. Second, we evaluate the full two-step framework—extrapolation followed by spatial interpolation (Section~\ref{sec:Spatial})—against the external validation dataset of hub-height measurements from operational wind farms.

\subsection{Accuracy on reanalysis data}
\label{sec:HeightResults}
We first benchmark the performance of our height extrapolation model against parametric baselines using the NEWA reanalysis dataset. The first baseline employs the power law with a constant shear exponent $\alpha = 1/7$, a common approximation for neutral atmospheric conditions over smooth terrain \citep{panofsky1984atmospheric}. The second baseline is a simplified, temporally varying $\alpha$ model following \cite{crippa2021temporal} (Section~\ref{sec:Height}), which incorporates diurnal harmonics but omits the temperature gradient to rely only on predictors available in real-time. These are compared against the GAM proposed in Section~\ref{sec:Height}.

A separate height-extrapolation model is fitted for each meteorological station using reanalysis wind speeds extracted from the nearest NEWA grid point and statistically downscaled using the Global Wind Atlas, as described in Section~\ref{sec:Preprocessing}. Predictions at any given time must rely solely on the 10\,m wind observation available at that location as models inputs. This setup ensures that model evaluation reflects the information constraints present in real-time applications.

All models are trained on 80\% of the NEWA reanalysis data and evaluated on the remaining 20\% holdout set. The root mean square error (RMSE) for each approach is presented in Table~\ref{tab:Reanalysis_acc}.

\begin{table}[h]
\centering
\begin{tabular}{|c|c|}
 \hline
    Model & RMSE (m/s) \\
 \hline
 Power Law ($\alpha=1/7$) & 2.92 \\
 Temporally Varying $\alpha$ & 1.53 \\
 GAM (Proposed) & 1.25 \\
 \hline
\end{tabular}
\caption{Comparison of height extrapolation accuracy on the reanalysis test set.}
\label{tab:Reanalysis_acc}
\end{table}

The results demonstrate that both temporally varying $\alpha$ and our proposed GAM offer substantial improvements over the constant power law. The proposed GAM provides a further, meaningful increase in accuracy, likely due to its ability to capture non-linear relationships between the 10\,m wind speed and the target height, as well as its incorporation of additional covariates like wind direction.

\subsection{Accuracy at wind farms}
We  evaluate the end-to-end performance of our two-step framework using an external test set comprising of hub-height wind speed measurements from seven operational wind farms (Section~\ref{Sec:WindFarms}). Using one year of 10-minute data, predictions are generated exclusively from publicly available datasets and the proposed modelling framework. For comparison, we also generate predictions from the ERA5 reanalysis dataset, a popular choice in renewable energy applications \cite{olauson2018era5,doddy2021reanalysis}. To construct a continuous time series from the hourly ERA5 data, wind speeds at 10\,m and 100\,m were extracted from the nearest ERA5 grid point to each wind farm location. The resulting hourly time series were then temporally interpolated to a 10-minute resolution. A site-specific power-law relationship was subsequently applied at each time point to estimate wind speeds at the actual turbine hub height.

Predicted wind speeds are evaluated against two validation observations defined in Section~\ref{Sec:WindFarms}: the maximum observed wind speed (a proxy for the unobstructed free-stream wind) and the farm average wind speed (which incorporates wake effects). As our model predicts the free-stream wind, its predictions are systematically higher than the farm average. To account for this and to facilitate comparisons, we apply simple empirical wake loss adjustments to predicted wind speed of 10\%, 15\%, and 20\%, reflecting typical values reported in the literature \cite{gonzalez2012wake, pryor2024wind, adaramola2011experimental}.

The results are summarised in Tables~\ref{tab:farm_acc_max} and \ref{tab:farm_acc_avg}, which evaluate model performance against two validation metrics. Table~\ref{tab:farm_acc_max} compares predicted wind speeds against the maximum observed wind speed at each farm, which serves as a proxy for the unobstructed wind speed. Under this metric, the proposed model achieves both a lower root mean squared error (RMSE) and a substantially smaller mean residual (bias) than the ERA5 benchmark, indicating improved accuracy and reduced systematic error in estimating unobstructed hub-height wind speeds. This suggests that combining data-driven height extrapolation with spatial interpolation better captures local wind regimes than relying on coarse-grid reanalysis products alone, even when evaluated at hub height.

Table~\ref{tab:farm_acc_avg} compares predictions against the farm-average wind speed, which reflects the combined effects of turbine wakes and local flow interactions. As the proposed framework predicts unobstructed wind speeds, the unadjusted model overestimates wind speeds and exhibits elevated RMSE when evaluated against farm average observations. However, after applying simple empirical wake-loss adjustments, the agreement improves substantially. In particular, the 15\% wake-loss adjustment yields the lowest RMSE and a mean residual close to zero, indicating stronger agreement with observed farm average wind speeds. Correlation coefficients are unchanged by the wake loss adjustment, as the adjustment corresponds to a linear scaling of the predicted wind speeds, which does not affect correlation. The sensitivity analysis across multiple wake-loss values demonstrates that modest wake corrections are sufficient to align raw predictions with farm level averages.

\begin{table}[h]
\centering
\begin{tabular}{|c|c|c|c|}
 \hline
    Model & RMSE (m/s) & Mean Bias (m/s) & Correlation\\
 \hline
 Proposed Model & 1.90 & -0.50 & 0.86\\
 ERA5 & 2.14 & 1.18 & 0.86\\
 \hline
\end{tabular}
\caption{Prediction accuracy aggregated across all seven wind farms, evaluated against the \textbf{maximum} observed wind speed at each farm (proxy for unobstructed free-stream wind speeds).}
\label{tab:farm_acc_max}
\end{table}

\begin{table}[h]
\centering
\begin{tabular}{|c|c|c|c|}
 \hline
    Model & RMSE (m/s) & Mean Bias (m/s) & Correlation\\
 \hline
 Proposed Model (Unadjusted) & 2.01 & -1.13 & 0.86 \\
 ERA5 & 1.72 & 0.38 & 0.86 \\
 Proposed Model (10\% Wake Loss) & 1.70 & -0.29 & 0.86\\
 \textbf{Proposed Model (15\% Wake Loss)} & \textbf{1.62} & \textbf{0.14} & 0.86\\
 Proposed Model (20\% Wake Loss) & 1.67 & 0.56 & 0.86\\
 \hline
\end{tabular}
\caption{Prediction accuracy aggregated across all seven wind farms, evaluated against the \textbf{average} observed wind speed at each farm.}
\label{tab:farm_acc_avg}
\end{table}

While predictive accuracy is comparable to or better than ERA5 under both validation metrics, the proposed framework additionally produces full predictive uncertainty, combining uncertainty from both height extrapolation and spatial interpolation.

\begin{figure}[!htbp]
    \centering
    \begin{subfigure}{0.48\textwidth}
        \centering
        \includegraphics[width=\textwidth]{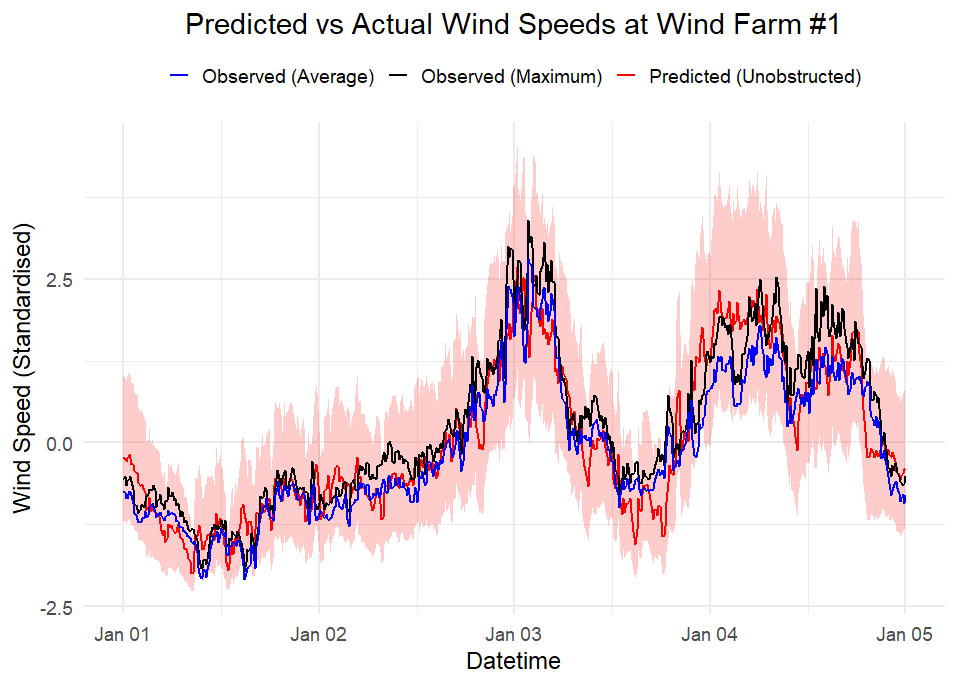}
        \subcaption{ (1) }
    \end{subfigure}
    \hfill
    \begin{subfigure}{0.48\textwidth}
        \centering
        \includegraphics[width=\textwidth]{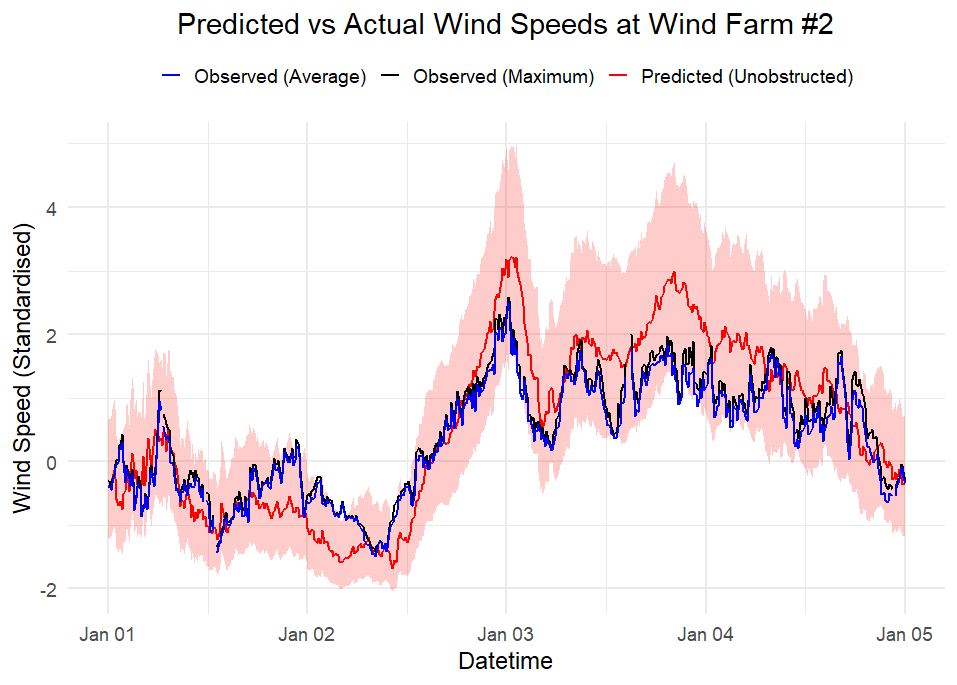}
        \subcaption{ (2) }
    \end{subfigure}

    \vspace{0.4cm}

    \begin{subfigure}{0.48\textwidth}
        \centering
        \includegraphics[width=\textwidth]{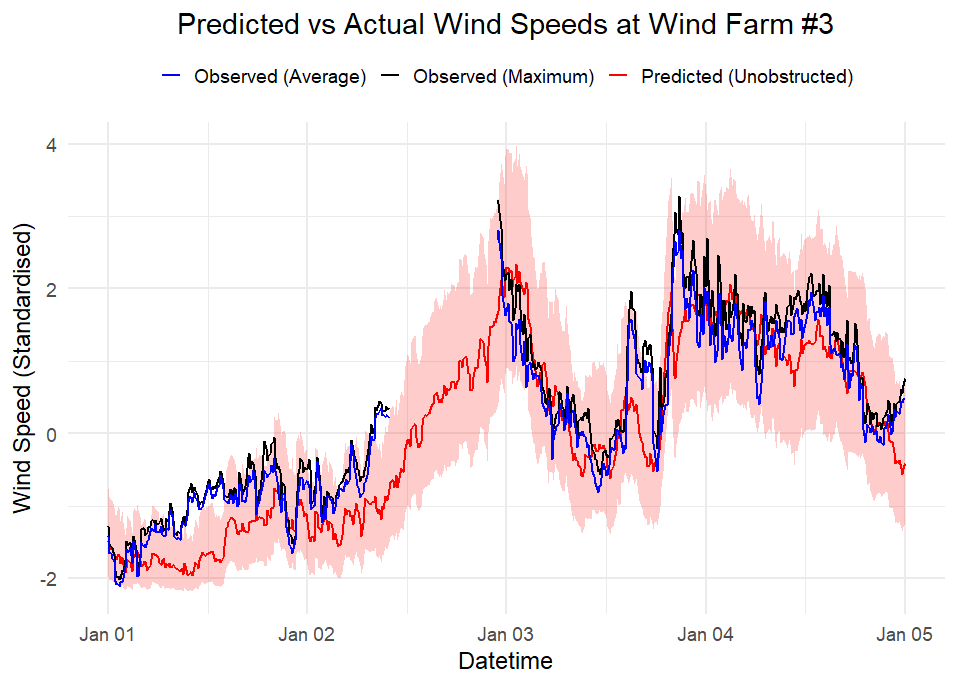}
        \subcaption{ (3) }
    \end{subfigure}
    \hfill
    \begin{subfigure}{0.48\textwidth}
        \centering
        \includegraphics[width=\textwidth]{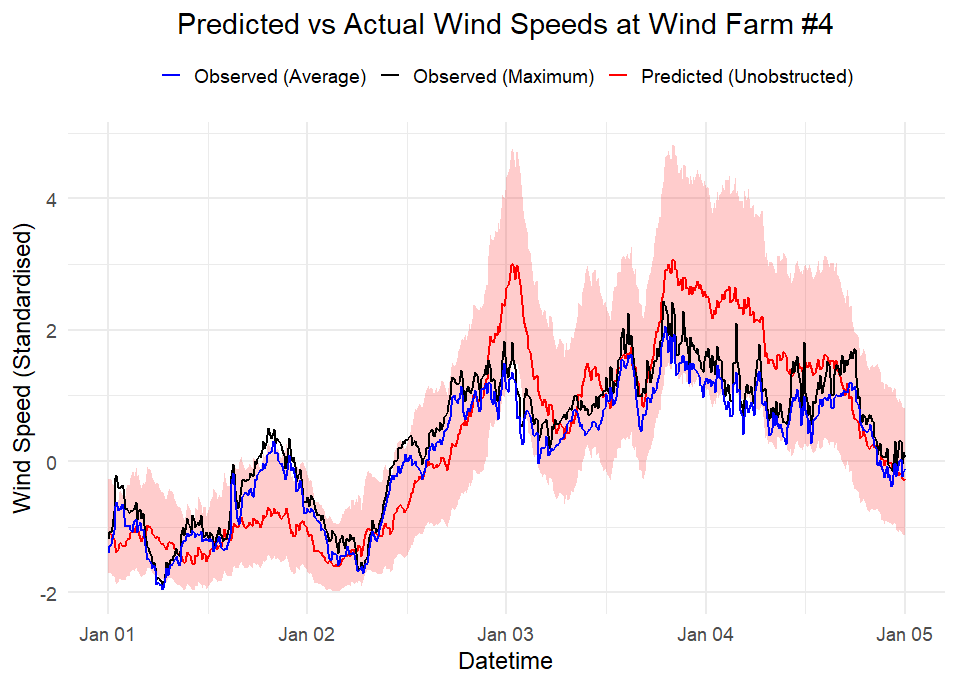}
        \subcaption{ (4)}
    \end{subfigure}

    \caption{\label{fig:FarmResults}Predicted and observed wind speeds at four wind farms over the first four days of the year (values have been standardised for data privacy). Shaded regions denote the 95\% prediction interval. In each panel, the average wind speed is shown in blue and the maximum wind speed in black. Periods with missing turbine observations are also visible.}
\end{figure}

Figure~\ref{fig:FarmResults} visualizes predicted hub-height wind speeds and 95\% prediction intervals at four sample wind farms over a five day period. The prediction closely follows the observed temporal trends, as well as rapid wind ramps. Observed wind speeds generally fall within the predicted uncertainty bounds, indicating that uncertainty estimates are well calibrated. Differences between the farm average and maximum observed wind speeds are also illustrated, consistent with the presence of wake effects within the farms.

Finally, we validate our uncertainty quantification by examining the empirical coverage of our prediction intervals (Table~\ref{tab:CoverageInterval}), which calculates what percentage of true values are contained within a given confidence interval. The 80\% and 95\% intervals show coverage of 85\% and 96.6\%, respectively, indicating that the estimated variance is slightly too high but well-calibrated.

\begin{table}[h]
\centering
\begin{tabular}{|c|c|}
 \hline
    Prediction Interval & Empirical Coverage \\
 \hline
 80\% & 85.0\% \\
 95\% & 96.6\% \\
 \hline
\end{tabular}
\caption{Empirical coverage of the prediction intervals at wind farms of our proposed model, demonstrating well-calibrated uncertainty quantification. These values are calculated using the maximum observed wind speed at all 7 validation sites, at each 10-minute time point over the full year. }
\label{tab:CoverageInterval}
\end{table}

\section{Discussion} 
\label{sec:Discussion}
This paper presented a two-step statistical framework to predict wind speeds at turbine hub heights using publicly available meteorological data from separate locations. By combining a non-parametric height extrapolation model with a spatial Gaussian Process, our approach produces high-resolution wind speed estimates at wind farm locations. When validated against seven onshore wind farms in Ireland, our model achieves comparable accuracy to ERA5 while being available in near real-time, highlighting its potential for operational use.

Given the open-access nature of the input data and prediction framework, this could be used as a tool alongside reanalysis data for preliminary site selection. Open-access wind resource tools can be important for smaller scale community energy groups who don't have the up front investment for extensive site surveys \citep{araveti2022wind}.
The framework provides a valuable tool for short-term wind monitoring and real-time decision-making. Unlike ERA5, which is available with a 5-day delay, our approach can deliver immediate estimates, enabling grid operators to adjust reserve capacity or optimize dispatch schedules \citep{wang2016value}. In addition, the framework generates continuous 10-minute resolution time series at any prospective wind farm location, enabling both real-time monitoring and retrospective analysis at spatial scales not directly available from reanalysis products. The provision of uncertainty estimates at each site further allows stakeholders to quantify the confidence in predictions and integrate probabilistic forecasts into operational planning, which is crucial for managing variability in wind generation.

Our model is sensitive to the choice of height extrapolation and the underlying reanalysis dataset. Non-parametric methods such as the Generalized Additive Model (GAM) used here allow flexibility in capturing local wind profiles, but performance may vary in regions with complex terrain, high forest cover, or limited data coverage. The use of reanalysis datasets inherently smooths local variability, which can under-represent extreme or highly localized wind events. Future work could explore the fusion of multiple reanalysis products or satellite-derived wind estimates to enhance model robustness \citep{gruber2022towards, dorenkamper2020making}.

While we focus on a two step approach, a joint multivariate model of wind speeds across multiple heights is conceivable. This strategy is impractical for our use case because (i) no hub-height data are available for conditioning the model during application, and (ii) hub heights vary between farms, complicating a unified approach. However for practitioners who have access to both meteorological station and hub-height observations, a multivariate approach, such as a coregionalisation model\citep{krainski2018chapter3}, or multi-output GP could be employed \citep{alvarez2008sparse}. 

Another limitation lies in the treatment of wake effects within wind farms. While we apply empirical reduction factors to account for speed losses at downwind turbines, more detailed modelling could incorporate turbine layouts and computational fluid dynamics (CFD) simulations \citep{gonzalez2012wake, pryor2024wind}. Such refinements would be particularly relevant when the model is applied to forecast farm-level energy output or turbine loads.

Beyond onshore wind, the methodology could be applied to offshore wind farms, where observational data are typically sparser, often limited to offshore weather buoys. It could also be applied to Advanced Scatterometer (ASCAT) satellite wind products, which provide near–real-time estimates of offshore 10,m wind speeds and have been validated in Irish waters \citep{remmers2019potential}. However, these products are available only at 10,m above sea level. Our two-step framework could therefore be used to extrapolate ASCAT derived surface winds to turbine hub heights and interpolate them to offshore wind farm locations. Accurate wind mapping at turbine heights could assist with short-term forecasting and resource assessment in emerging offshore markets, such as Ireland’s planned 5 GW offshore wind capacity. Combining our statistical framework with numerical weather prediction (NWP) outputs may further improve sub-hourly forecasting, addressing a key limitation of traditional NWP models, which typically provide data at 10\,m heights \citep{olauson2018era5, gualtieri2019comprehensive}.

From a practical perspective, this framework could reduce reliance on expensive on-site wind measurements for preliminary assessments. Traditional wind resource assessments require extended campaigns, often costing tens of thousands of euros per site \citep{TeagascWind}, whereas our approach leverages open-access datasets to provide rapid, high-resolution estimates. This has clear implications for project development, investment decisions, and grid management, particularly when considering multiple candidate sites or monitoring operating wind farms in real time.

In summary, our two-step approach enables high-resolution, real-time predictions of turbine-height wind speeds from publicly available data, with quantified uncertainty. While the method performs comparably to ERA5 in an Irish context, future work should focus on improving wake effect modelling, exploring multi-source data fusion, and extending the framework to offshore applications to maximize its practical impact.

\section*{ Declaration of generative AI and AI-assisted technologies in the writing process}
During the preparation of this work the author(s) used GPT4 of openAI in order to improve
the language and structure. After using this tool/service, the author(s) reviewed and edited the
content as needed and take(s) full responsibility for the content of the publication.

\section*{Acknowledgements}
This publication has emanated from research supported in part by a grant from Taighde Éireann – Research Ireland under Grant number 18/CRT/6049. This publication has emanated from research conducted with the financial support of
the EU Commission Recovery and Resilience Facility under the Research Ireland
Energy Innovation Challenge Grant Number 22/NCF/EI/11162G. This work was partially supported by the “PHC ULYSSES” program (project number:  50252NC ), funded by the French Ministry for Europe and Foreign Affairs, the French Ministry for Higher Education and Research, Taighde Eireann – Research Ireland.
The French authors also acknowledge the financial support of the Chair Geolearning, funded by ANDRA, BNP Paribas, CCR and the SCOR Foundation for Science. 
For the purpose of Open Access, the author has applied a CC BY public copyright licence to any Author Accepted Manuscript version arising from this submission.

Data obtained from the Global Wind Atlas version 4.0, a free, web-based application developed, owned and operated by the Technical University of Denmark (DTU). The Global Wind Atlas version 4.0 is released in partnership with the World Bank Group, utilizing data provided by Vortex, using funding provided by the Energy Sector Management Assistance Program (ESMAP). For additional information: \href{https://globalwindatlas.info}{https://globalwindatlas.info}

Data is also obtained from the New European Wind Atlas, a free, web-based application developed, owned and operated by the NEWA Consortium. For additional information see \href{https://map.neweuropeanwindatlas.eu/www.neweuropeanwindatlas.eu}{www.neweuropeanwindatlas.eu}

The authors thank Alejandro Fernandez and Greencoat Renewables for providing the wind farm data used for validation in this study.








\bibliographystyle{elsarticle-num}
\bibliography{elsarticle_bib}


\end{document}